# Cavity Continuum

FAN CHENG[1], VLADIMIR SHUVAYEV[2], MARK DOUVIDZON[3], LEV DEYCH[4] AND TAL CARMON[1,*]

[1]*School of Electrical Engineering, Tel Aviv University, Tel Aviv 6997801, Israel*
[2]*Physics Department, Queens College of CUNY, Flushing, Queens, New York 11367, USA*
[3]*Solid State Institute, Technion-Israel Institute of Technology, Haifa 3200003, Israel*
[4]*The Graduate Center of CUNY, 365 5th Ave., New York, New York 10016, USA*

*[total@tauex.tau.ac.il](mailto:total@tauex.tau.ac.il)*

**ABSTRACT**

**We experimentally demonstrate and numerically analyze large arrays of whispering gallery resonators. Using fluorescent mapping, we measure the spatial distribution of the cavity-ensemble's resonances, revealing that light reaches distant resonators in various ways, including while passing through dark gaps, resonator groups, or resonator lines. Energy spatially decays exponentially in the cavities. Our practically infinite periodic array of resonators, with a quality factor [Q] exceeding $10^7$, might impact a new type of photonic ensembles for nonlinear optics and lasers using our cavity continuum that is distributed, while having high-Q resonators as unit cells.**

## 1. INTRODUCTION

High Q optical resonators [1,2] supported microlasers [3–8], nonlinear optics [9,10], and experiments in cavity optomechanics [11] and optocapillaries [12]. As one can see in Fig. 1, optical resonators were extended to an array of a few and then several resonators. Here we mass-produce microdroplet resonators with optical Q exceeding $10^7$ and cascade them into a cavity continuum of practically unlimited size. In detail, coupling between a pair of resonators permitted enhanced sensing using exceptional points [13–17]. Cascading three cavities allowed flattened bandpass filters [18]. Then, going to several resonators exhibited families of modes, repeating every free spectral range [19,20]. Clusters of 5 microparticles [21], each with Q near 1000, allowed biosensing via the ensemble's spectral fingerprint. Massive 3-D arrays of dielectric microspheres with many resonators and Q = ~1000 were exhibiting anomalously high transmission at the WGM peak wavelengths [22]. On another extreme, random lasers [23–26] represent a structure where coincident multi-scattering, e.g., by porous media, provides feedback, while each scatterer is a non-resonant body or, in different words, with a very low Q, typically near 1. As for the arrangement of resonators [27,28], cavities can be coupled in a linear array [29–31,19], branched structures [32,33], 3D structures [22,34,35], and a close-packed layer that we present here.

Here we introduce and study a continuous resonator-array. This type of cavity continuum might support detection via the ensemble's spatial and spectral fingerprint [21] at higher Q, benefiting better sensitivity to analytes. This cavity ensemble might also permit a new type of random lasers [23–26], and delay lines [28], or enable phase matching for nonlinear effects via a denser power spectrum [7]. In a broader context, one can describe our cavity continuum as a new type of photonic ensembles where each unit cell is a spherical resonator with Q exceeding $10^7$.

Droplets were functioning as optical cavities with high Q [12,36–40] and activated as Raman [41], Brillouin [42], and water-ware lasers [43]. Coupling between resonators requires nanometer-scale control over the distances between resonators, which is challenging. Nevertheless, droplet resonators are relatively easy to mass-produce [44–47] and self-arrange into a periodic array of almost touching droplets. Despite that, coupling many high-Q microresonators into a continuous array was rarely studied, not even theoretically [48–50]. Following the recent demonstration of optical coupling between several droplet resonators [19], we extend this ensemble here to a cavity continuum that is practically unlimited in size. The current practical limit on the size of our array is near 400, limited by the sensitivity of our camera.

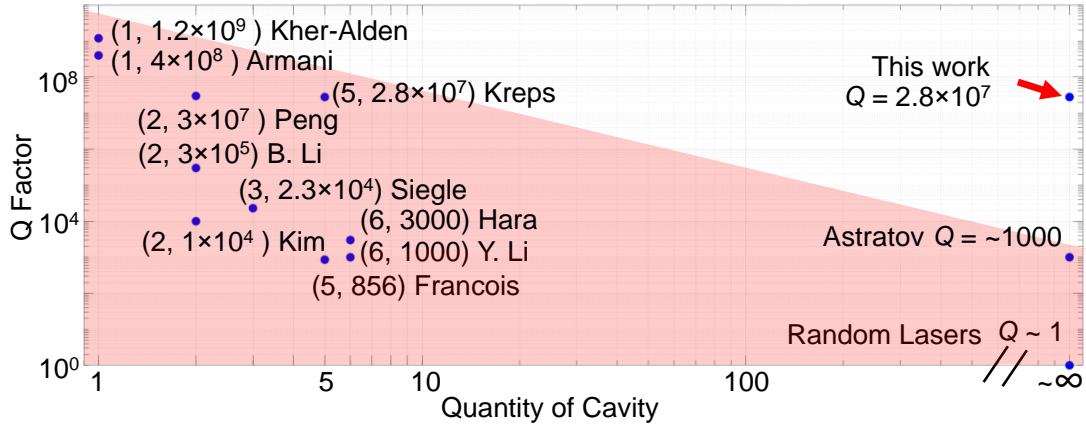

**Fig. 1.** Number of micro resonators vs Q factor [51,29,21,16,52–54,33,39,19,22] represents a trade-off between the quality factor and the number of resonators. Our work on the top right uniquely combines high optical Q with a large number of resonators.

We examine the optical modes of the resonator ensemble using the fluorescent mode mapping technique. In detail, doping the liquid surrounding the droplets with a fluorescent material makes this region glow in places where the droplet's mode evanescently extends to its surroundings. We can therefore perform a wavelength scanning of the input light while taking a movie of the fluorescent emission. The micrographs representing the movie's frames reveals the resonance's spatial structure, while the time variation of the emission from each resonator represents its optical absorption. In this manner, the fluorescent-based technique maps resonances both spatially and spectrally. It is helpful that fluorescent dopants convert some of the coherent resonance light into incoherent radiation emitted in all directions. This is because, unlike coherent light, fluorescent emission does not suffer speckles and is therefore benefitting higher quality images. Such fluorescence was used to map the modes of spherical resonators [3,4] as well as for measuring stopped light in such spheres [55]. In toroids, level-crossing phenomena were fluorescently photographed [56], and in liquids droplets, fluorescent mapping resolved the structure of droplets whispering gallery modes [57], as well as of the fluidic vortices [58] they generate. In a similar manner, the optical modes of a water fiber were mapped [59]. Later on, modes of a few-droplet ensemble [19] were fluorescently mapped.

## 2. EXPERIMENTAL SETUP

In our experimental setup, as depicted in Fig. 2, we employ a 780 nm tunable laser (New Focus, TLB-6712-P) as a light source. We use a curved tapered fiber to couple light into the droplets. A straight tapered fiber is fabricated by heating a single-mode fiber (Thorlabs, 780HP) with a hydrogen flame while drawing it with two translation stages. To make this taper bent [60], for the purpose of accessing only several droplets, we release the straight tapered fiber to generate a curved variant and heat it again for annealing, producing a mechanically stable fiber [61]. The curved tapered fiber is rotated by two fiber rotators (Thorlabs, HFR007) until the curved part faces downwards. This can allow the thinnest part to touch a few droplets while the tapered regions are far from obstacles. Droplet generation is explained in APPENDIX A.

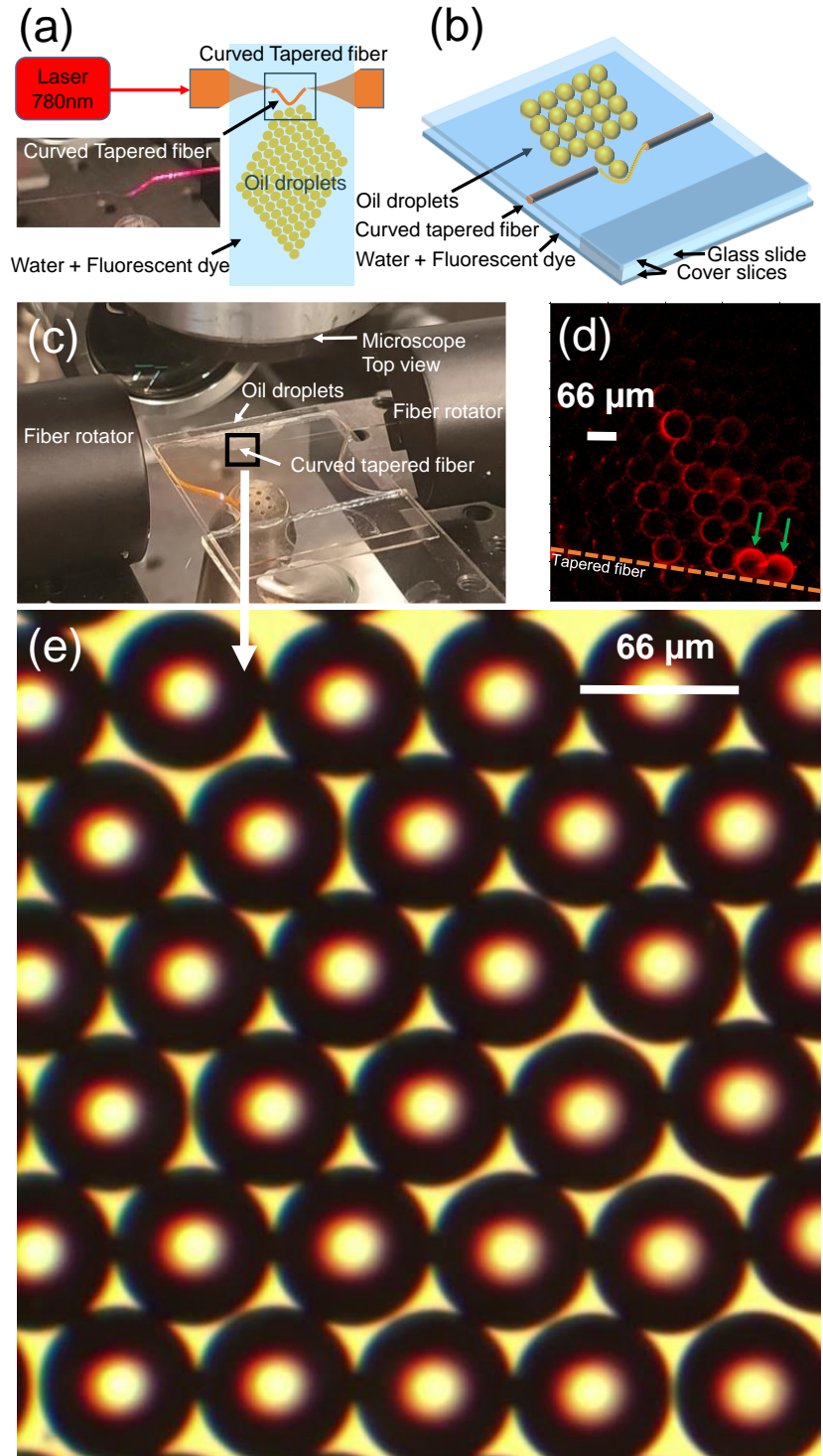

**Fig. 2.** Coupling setup. (a) A tunable laser (780 nm) serves as the laser source and a curved tapered fiber is used to couple oil droplets. (b) A chamber is assembled by two cover slices and a glass slide (thickness: 1 mm). The chamber is filled with deionized water and fluorescent dye. The oil droplets are placed in the water environment to maintain their spherical shape. The fluorescent dye is used for mapping the mode. (c) A vertical microscope is used to observe the coupled droplets. Two fiber rotators are used for the fabrication of the curved tapered fiber. (d) A top view image taken by the vertical microscope and a near IR camera reveals coupled droplets with a highlighted circumference, with the orange dashed line representing the curved tapered fiber. The coupled resonators are marked by green arrows. (e) The clusters are spatially oriented in a hexagonal lattice configuration [62], where the standard deviation in droplet diameter is 0.5 µm and the standard deviation in resonator-to-resonator distance is 1.5 µm. Each droplet is coupled to other 6 droplets with 84% success, as indicated by occasional gaps between droplets.

The curved tapered fiber is positioned in a chamber constructed from two cover slices and a glass slide (thickness: 1 mm), which is filled with deionized water and fluorescent dye (American Dye Source Inc., ADS780WS, 3 µmol/L [57], absorption peak at 780 nm, emission peak at 813 nm). The chamber can maintain a relatively stable aquatic environment due to the water's surface tension, preventing leakage around the edges. The fluorescent dye is dissolved into the deionized water to be later used for mapping the optical mode. To distinctly observe the fluorescence and eliminate the scattering of the laser, a long pass filter (Thorlabs, FELH0800, cut-on wavelength: 800 nm) is inserted between a vertical microscope and a near-infrared [NIR] camera (Lumenera, Infinity 3S) (APPENDIX B).

A pipette is employed to transfer the oil droplets to the edge of the chamber, where the droplets spread to form two-dimensional ensembles of droplets. The chamber is fixed onto a translation stage (PI, microtranslation stages: M-105; nanocube nanopositioner: P-611.3), facilitating the control of the relative positioning between the droplets and the tapered fiber. The droplets are brought close to the thinnest part of the fiber while ensuring the fiber is situated at the same level as the equatorial plane of the droplets. This alignment allows the optical modes of the multiple mutually coupled droplets to maintain a shared plane that runs parallel to the base of the chamber, thereby accomplishing the coupling among multiple droplets. 2 droplets are coupled by the tapered fiber (the coupled resonators are marked in Fig. 2d by green arrows).

## 3. EXPERIMENTAL RESULTS

Droplets of uniform size are self-arranged in a periodic array on a 2D plane while interconnected at their equatorial planes. We start by fluorescently mapping a cavity continuum (Fig. 3) by measuring its spectral transmission response in combination with the spatial mode structure for each of the absorption peaks. The fluorescent intensity as a function of time (Fig. 3a) stands for the spectral absorption of the cavity continuum and exhibits several absorption lines with $Q = (4.3 \pm 1.1) \times 10^5$. We note that the quality factor of the bare resonators (without the fluorescent ink) exhibits Q exceeding $10^7$, as indicated by the resonance linewidth. Micrographs corresponding to individual absorption peaks are shown in Fig. 3b (M1-M9) and represent the spatial distribution of light for these absorption peaks (Fig. 3a). One of the micrographs (M6) (Fig. 3c) is enlarged to resolve level crossing between modes is frequently, as indicated by a flower-shaped resonance – representing optical interference between two modes having similar resonance frequencies but a different number of wavelengths along the equator line [55–57]. For example, resonator (-4, 2) in Fig. 3c has ten maxima along its circumference arranged in a shape that resembles petals. Given the parameters of the resonator, these 10 circumferential fringes are most likely originating from interference between a mode with 395 waves resonating along the circumference and a mode with 405 circumferential waves. These modes are resonating at the same frequency, to within the resonance full width at half max. Though modes of different principal quantum numbers, as in Fig. 3c resonator (-4, 2), are expected to resonate at different frequencies, other eigen indices of the modes (e.g., its radial index) can compensate for this spectral gap. As one can see in Fig. 3b, each of the M1-M9 modes has a unique spatial signature.

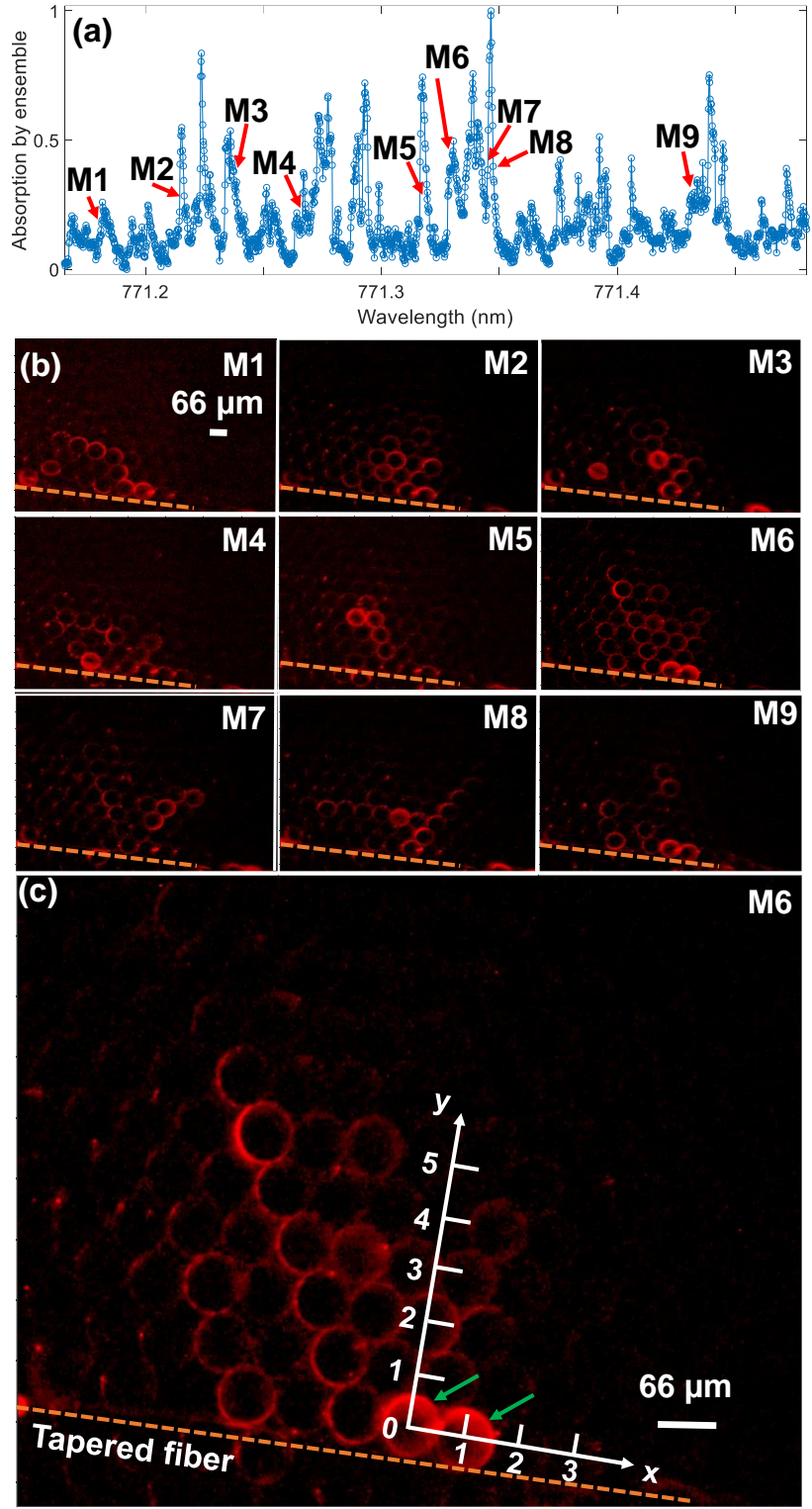

**Fig. 3.** Fluorescent mode mapping of a cavity continuum. (a) Spectral mapping of the ensemble's absorption lines as indicated by its total fluorescent emission. Several absorption lines are chosen to be accompanied by their spatial mapping (b) as correspondingly indicated in micrographs M1-M9. (c) Ensemble's mode structure with level-crossing events in resonators (-2.5, 1), (-4, 2), (-2, 2), (-3, 4) and (-4, 6)**.** The coupled resonators are marked by green arrows. The orange dashed line represents the curved tapered fiber. The wavelength scan took 30 seconds (see **Visualization 1**).

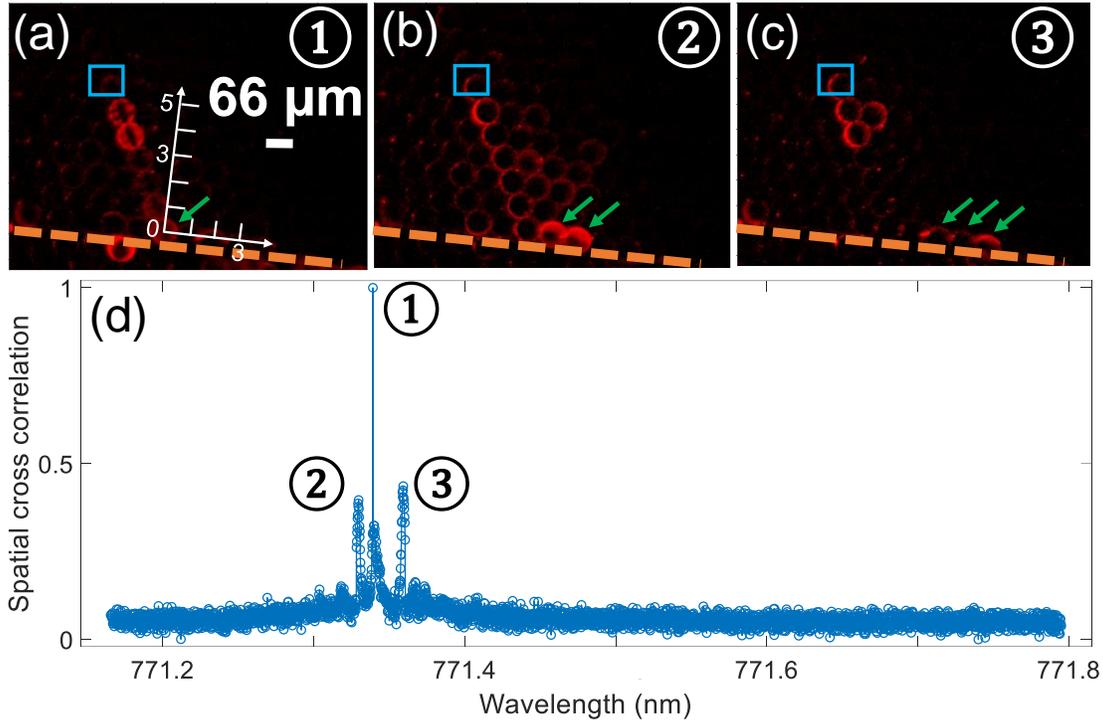

**Fig. 4.** Ways for light to reach a resonator (marked by the blue rectangle, coordinates (-3, 5.5)) (a) through a row of resonators (b) through a group of resonators (c) through dark resonators. (d) correlation between modes. The orange dashed line represents the curved tapered fiber. The wavelength scan took 60 seconds (see **Visualization 1**). The coupled resonators are marked by green arrows.

    We will now measure the spectral properties of an individual resonator in parallel with photographing the spatial distribution of light in the ensemble while this specific cavity is resonating (Fig. 4). This will tell us about the various ways light can reach a specific region of the ensemble. To do so objectively, we photograph a picture at a spectral wavelength supporting a resonance and mark a region of interest [ROI] (Fig. 4a-c blue rectangle) that includes a single resonator. We check for the autocorrelation of this ROI photograph with each of the movie frames (APPENDIX C). The strongest autocorrelation peaks indicate this specific resonator's excitation, as shown in Fig. 4d. Fig. 4a-c shows the spatial distribution of light corresponding to these correlated absorption peaks. As one can see in Fig. 4a-c, light can reach the resonator through a row of resonators, through a group of resonators, and, maybe surprisingly, through dark resonators. The quality factor of the individual resonator is $(4.5 \pm 1.5) \times 10^5$, which is slightly higher than the quality of the ensemble (Fig. 3). There exist two possible mechanisms of the optical transports in the arrays of microspheres: evanescent coupling of the whispering gallery modes, which might result in formation of the collective modes of the arrays [29,48,49] and so-called photonic nanojets [63–65]. The latter is a nonresonant phenomenon resulting from focusing of an incident light beam on the shadow side of a microsphere. Authors of Ref. [66–69] showed that such photonic jets can propagate along a long chain of microspheres. There are strong indications that the optical transport observed in our experiments is due to the evanescent coupling rather than nanojets. Indeed, Nanojets are usually excited by a free propagating light beam or as in case of Ref. [66] by a broadband fluorescence from an adjacent sphere, while the excitation mechanism in our experiments involve a tapered fiber, which selectively excites only a small number of WGMs, and thus cannot excite photonic jets. The main impediment to the optical transport due to the evanescent coupling is the variation of the sizes of the spheres resulting in the frequency mismatch between resonators. In our experiments, however, the size dispersion of the droplets is less than 1%, which according to Ref. [50] would have resulted in a linear chain of resonators to the localization length of the order of tens of resonators. Moreover, the evanescent transport in our system can be further enhanced due to the relatively large size of the droplets. Indeed, resonators with larger diameters are characterized by a smaller free space region. As a result, WGM with different orbital numbers can spectrally overlap in

spheres with different diameters, providing an efficient coupling even between size-mismatched droplets.

To support our experimental findings, we also carried out numerical simulations of an array of resonators. Since simulating numerous coupled three-dimensional droplet resonators using finite element method encounters significant computational difficulties, we limited our simulations to an array of two-dimensional disk resonators shown in Fig. 5 with a single resonator coupled to the waveguide. Simulations were performed within the regions close to the surface of resonators inside and outside, and the widths of the regions were chosen to keep the light propagation unaffected by the boundaries. We considered an array of 25 closely packed disks arranged in the form of a parallelogram shown in Fig. 5 with the main diagonal at 45° angle with the fiber. The exact orientation of this structure is not significant and is chosen in order to avoid direct excitation by the fiber of more than one resonator. We also did not incorporate in our simulation the variations in the sizes of the disks because this effect was assumed to be small as explained above. This approach is justified by the fact that the goal of the simulations is to demonstrate qualitatively that various propagating paths of light observed in our experiments are real physical phenomena rather than to achieve quantitative agreement between the simulations and the experiment. The later goal would have been unattainable even if we had managed to simulate the actual three-dimensional structures because the exact positions of the droplets in the experiment are not known, while the results of the simulations are sensitive to these parameters. The results of the simulations shown in Fig. 6 reproduce qualitatively various propagating paths of light observed experimentally.

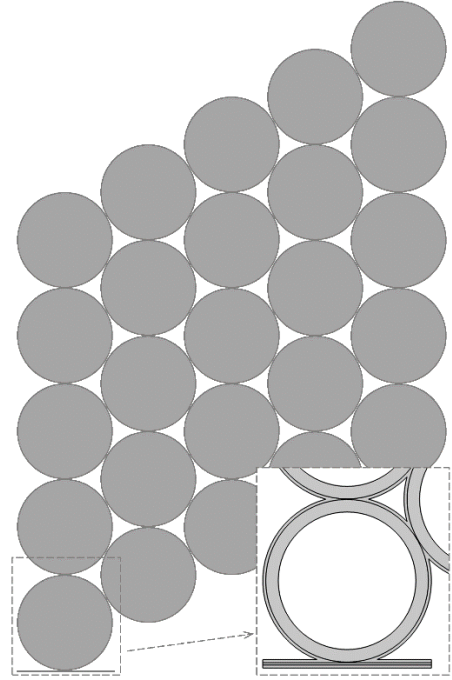

**Fig. 5.** Two-dimensional array of disks with a single disk excited by the waveguide (left-bottom corner of the figure). The inset shows domains where numerical simulations were performed.

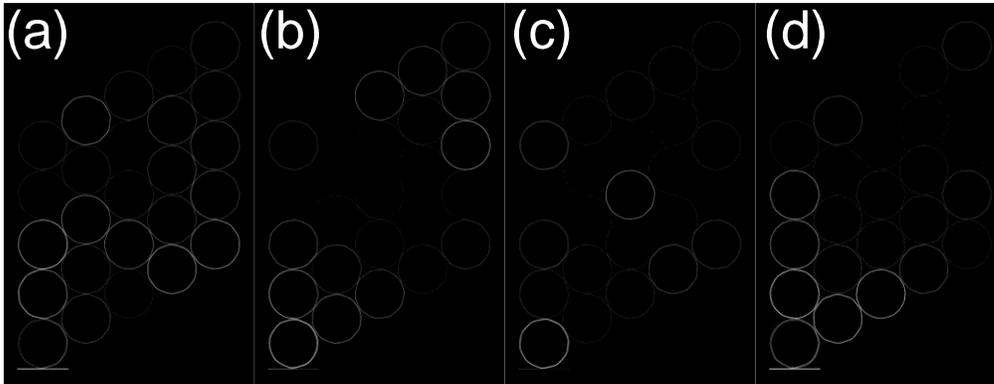

**Fig. 6.** Various propagating paths of light for different excitation wavelengths. The contrast and brightness of the images were modified to enhance the details of the light distribution.

One of the most interesting configurations observed in real and numerical experiments is the one in which bright resonance is separated by significant dark regions (Fig. 6b and Fig. 4c). Similarly, one can see a calculated "V" shaped resonance structure (Fig. 6d) and a photograph of its experimentally measured parallel (Fig. 8a).

Such a distribution of the intensity can be explained by destructive interference of light taking place within the dark regions between illuminated resonators. This phenomenon can be compared with the distribution of light intensity in a linear chain of resonators studied theoretically in Ref. [30] and experimentally in Ref. [32,33], where certain collective modes of the resonator chains would be characterized by vanishing light intensity inside individual resonators separating brightly lit neighbors.

We will now measure the energy decay as a function of distance from the tapered fiber coupler. We do so by first averaging the movie's frames while scanning the laser wavelength from 771.0 to 771.6 nm (Fig. 7a). Then we average over all the lines perpendicular to the tapered fiber and fit the intensity decay to an exponential function (Fig. 7b). We see that the decay is exponential, with a characteristic decay distance of 169 microns, meaning that power drops to 37% when reaching this distance from the taper. The number of cavities is limited by the energy decay (Fig. 7) and by camera sensitivity. The exponential decay of the averaged intensity is likely a result of combination of light attenuation and might be related to Anderson localization of light in this system. While it is not the goal of this paper to study the light localization phenomena, we would like to point out that the coupling between well confined whispering gallery modes is most naturally described in terms of a tight-binding model with nearest-neighbor interaction [2] and both diagonal and non-diagonal disorder. The Anderson localization phenomenon in such models is well established [70]. An alternative approach based on the percolation theory proposed in Ref. [22] can also be used to explain the origin of various paths of light propagation in our system, while our numerical simulation assists with the spatial dependence of the averaged intensity.

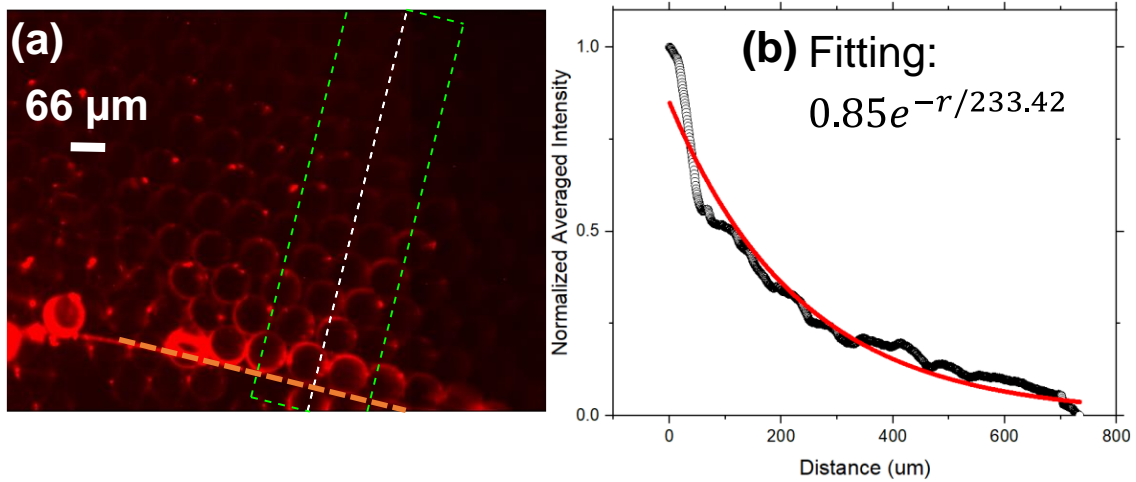

**Fig. 7.** Energy decay as a function of distance from the tapered fiber. (a) Average of the movie's frames while the input light is scanning from 771.0 to 771.6 nm. (b) Average over between lines perpendicular to the tapered fiber together with an exponential fit. The orange dashed line represents the curved tapered fiber (see **Visualization 2**).

As for the size of the cavity continuum, we can now measure -- in our experiment, we observed resonances that are excited at a 16-resonator distance from our optical coupler (Fig. 8). Therefore, we are capable of monitoring cavity continuum in a $\pi \times 16^2/2 = 402$ in a typical experiment. We believe that a cooled CCD and syringe pumps with more stable flow rate for smaller droplets' size variations, to be implemented in future experiments, will considerably increase the size of the accessible cavity continuum.

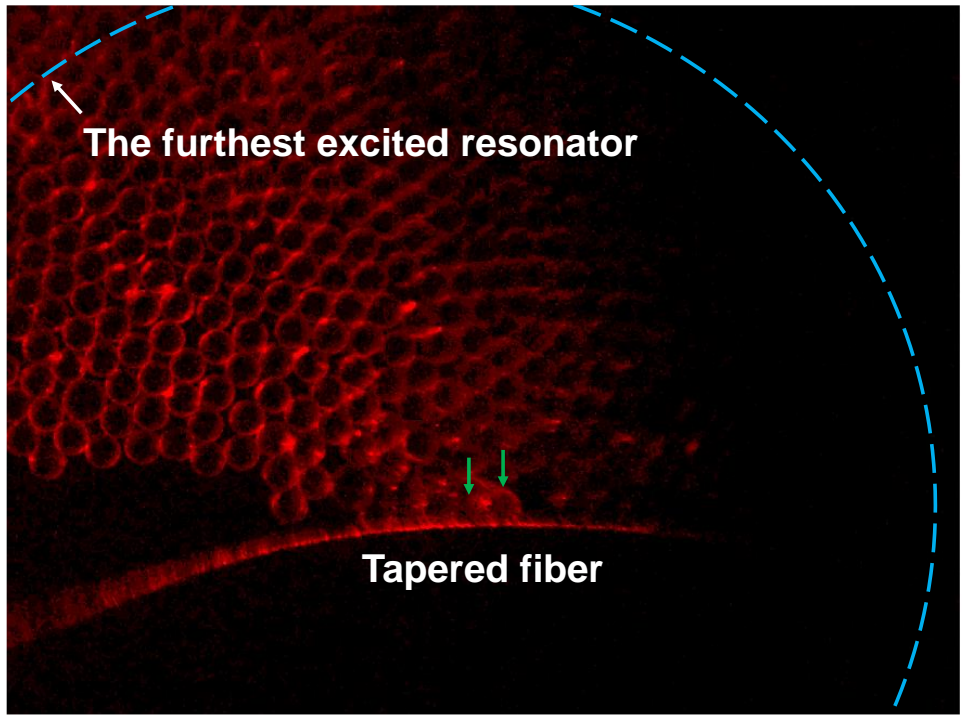

**Fig. 8. Cavity number.** We could measure a cavity continuum containing 402 resonators. The blue line represents the longest distance from the coupling point, where resonances, in the form of a ring of light appearing upon wavelength scan, were still visible. This blue region contains 402 resonators. This number of cavities is currently limited by our camera sensitivity and droplets' size variations [71]. The green arrows near the tapered fiber describe the resonators coupled to the tapered fiber.

Lastly, we depicted photographs describing various spatial structures in the ensemble as well as its individual cavity resonances. Fig. 9a demonstrates that droplet 1, being in direct contact with the fiber, is coupled to two linear arrays of droplets arranged in a "V" shape. Another coupling phenomenon is observed whereby a single droplet exhibits a resonant structure constructed of rings tilted one in respect to the other (Fig. 9b). Furthermore, as photographed in Fig. 9b and a complimentary movie (see Visualization 2), a continuous tilting cycle is observed of the droplet's mode while we scan the wavelength.

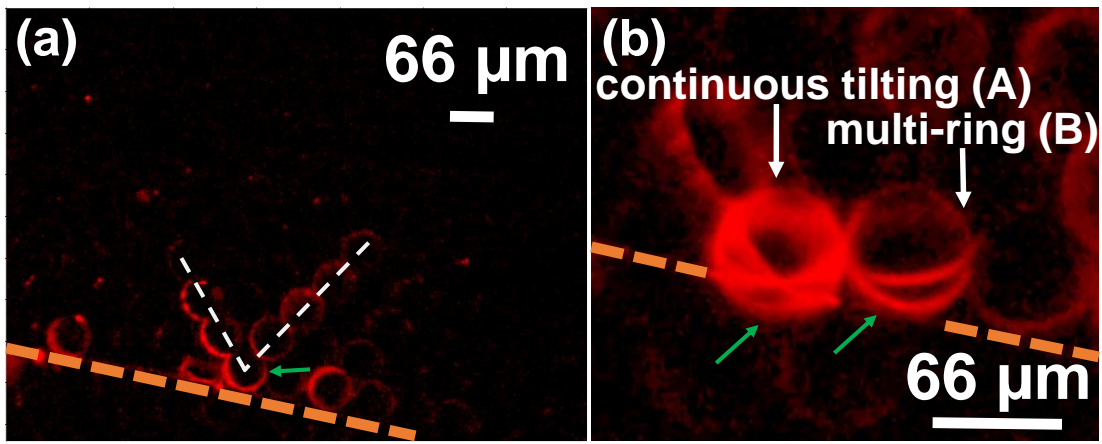

**Fig. 9.** A variety of mode structures. (a) A collective V-shaped resonance (b) Individual resonances with multi-ring shapes and continuous tilting during a wavelength scan (see **Visualization 2**). The orange dashed line represents the curved tapered fiber. The coupled resonators are marked by green arrows.

## 4. CONCLUSIONS

This study optically couples a large number of liquid resonators on a 2D plane via a curved tapered fiber. This new entity is periodically continuous, as typical to crystals, with high Q resonators functioning as unit cells. We believe this study will open a window into a new type of photonic ensembles where many droplets will be arranged in lines, circles, polygons, arbitrary shapes, and various 2D and 3D arrays to investigate new types of photonic continuums.

**APPENDIX A: GENERATION OF LIQUID RESONATORS**

In the construction of liquid resonators, immersion oil is employed (Sigma, model 56822, n = 1.516, 1.025 g/mL). This high-refractive index oil was selected to provide a large index contrast relative to the surrounding water (n = 1.333) to support whispering gallery resonances while suffering minimal optical losses due to radiation [72]. Additionally, this specific oil was chosen for providing quality factors exceeding $10^7$ [19]. The liquid resonators are generated using the X-junction of a microfluidic chip (Dolomite, Part number 3000158). As illustrated in Fig. 10, flow rates of oil and deionized water are regulated by two syringe pumps (Chemyx Inc, Fusion 100) (oil: 1 mL syringe, 100 µL/min; water: 20 mL syringe, 100 µL/min). The procedure to generate droplets commences with the initiation of the water syringe pump to ensure the entire chip becomes submerged with water, followed by the activation of the oil syringe pump. Inside the chip, the oil phase is subjected to pressure from two water channels, resulting in the generation of oil-in-water droplets. Once a stable production of droplets is confirmed, droplets are subsequently collected in a test tube. This control over flow rates ensures the creation of uniformly sized droplets with a diameter of 66.3 ± 0.5 µm where accuracy is limited by the resolution of our microscope. Typically, hundredths of thousandths of droplets are generated during a single batch. These droplets, possessing a density greater than that of water, eventually settle at the bottom of the test tube, facilitating the formation of an array where droplet-to-droplet contact is enabled.

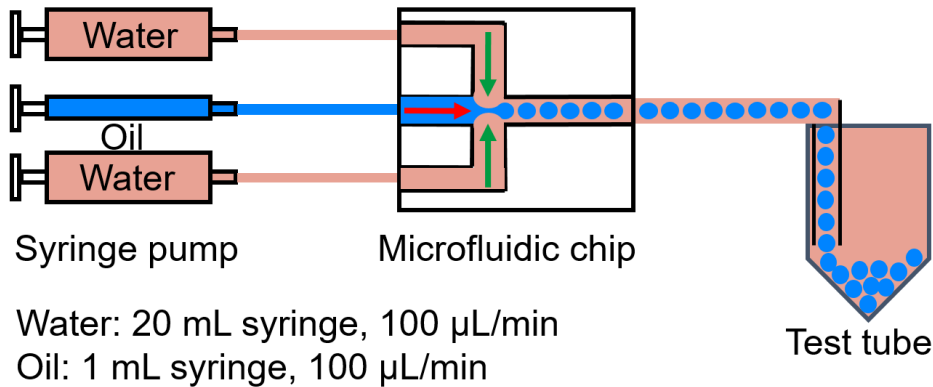

**Fig. 10.** Droplets generation: Schematic diagram of the system generating oil-in-water droplets. Two 20 mL syringes are charged with deionized water, while a 1 mL syringe is filled with immersion oil. The three syringes are propelled at an identical rate of 100 µL/min. The water acts to compress the oil at the X-junction within a microfluidic chip, thus yielding homogeneous oil-in-water droplets. These collected droplets are then gathered at the base of a test tube due to their higher density compared with water.

**APPENDIX B: FLUORESCENT MATERIAL PROPERTIES**

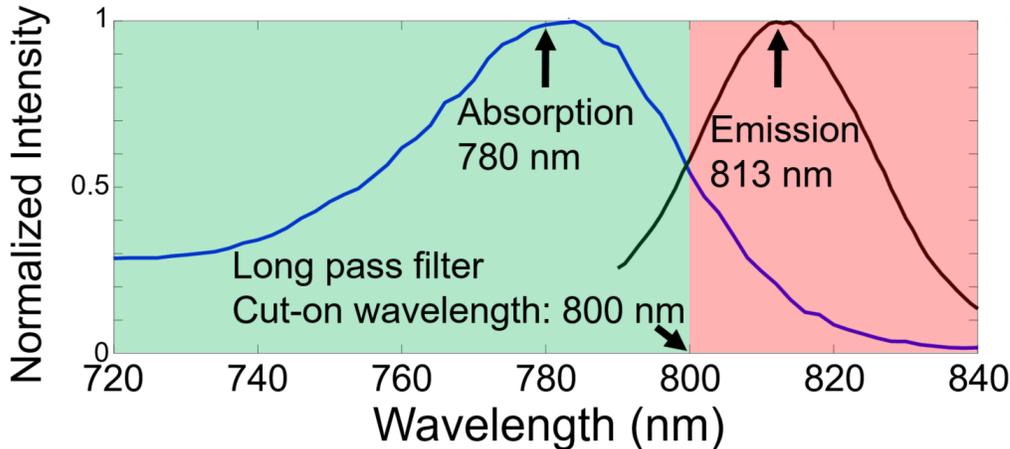

**Fig. 11.** Fluorescent material properties. (Left) absorption of the fluorescent material is at the optical resonance wavelengths that we study here. (Right) Fluorescent emission at a wavelength longer than the long pass filter cut on, where the NIR CCD camera is still sensitive.

**APPENDIX C: AUTOCORRELATION**

In Fig. 4d, the normalized intensity of each pixel in the blue rectangle in Fig. 4a is multiplied by that of the corresponding pixels in another frame, and then all multiplications are summed to provide the correlation of this specific photograph with the test-case photograph where resonance 7 is excited. This procedure is performed on all of the frames in the movie that is taken during the wavelength scan. The highest peak therefore stands for the modes where resonator number 7 is excited.

**Funding.** United States–Israel Binational Science Foundation (NSF-BSF) (2020683); Israeli Science Foundation (537/20).

**Acknowledgments.** We thank Amirreza Ghaznavi, Jie Xu, Yue Wu, Gilad Yossifon, Stanislav Kreps, Shenglong Liao for their useful technical discussions. F.C. performed the experiments. V.S. performed the theoretical analysis. M.D. and L.D. supervised the work. T.C. supervised all aspects of the work.

**Disclosures.** The authors declare that there are no conflicts of interest related to this article.

**Data Availability.** Data underlying the results presented in this paper are not publicly available at this time but may be obtained from the authors upon reasonable request.